\documentclass[prl,twocolumn,showpacs,aps,floatfix,fleqn]{revtex4}

\usepackage{amsmath}
\usepackage{amstext}
\usepackage{epsfig}
\usepackage{graphicx}
\usepackage{dcolumn}
\usepackage{rotating}
\usepackage{draftcopy}
\usepackage{longtable}
\usepackage{caption2}
\usepackage{flafter}
\usepackage[below]{placeins}

\begin{document}

\bibliographystyle{apsrev}


\title{Land\'e $g$ factors and orbital momentum quenching in semiconductor quantum dots}

\author{Craig E. Pryor}
\author{ Michael E. Flatt\'e}

\affiliation{Department of Physics and Astronomy and Optical Science and Technology Center, \\ University of Iowa, Iowa City, Iowa 52242}

\date{\today}

\begin{abstract}
We show that the electron and hole Land\'e $g$ factors in self-assembled III-V quantum dots have a rich structure intermediate between that expected for paramagnetic atomic impurities and for bulk semiconductors.  Strain, dot geometry, and confinement energy significantly modify the effective $g$ factors of the semiconductor material from which the dot and barrier are constructed, yet these effects  are insufficient to explain our results. We find that the quantization of the quantum dot electronic states  further quenches the orbital angular momentum of the dot states, pushing the electron $g$ factor towards $2$, even when all the semiconductor constituents of the dot have negative $g$ factors.  This leads to trends in the dot's electron $g$ factors that are the opposite of those expected from the effective $g$ factors of the dot and barrier material. Both electron and hole $g$ factors are strongly dependent on the magnetic field orientation; hole $g$ factors for InAs/GaAs quatum dots have large positive values  along the growth direction and small negative values in-plane. The approximate shape of a quantum dot can be determined from measurements of this  $g$ factor asymmetry.
\end{abstract}

\pacs{}

\maketitle

An individual electron or hole spin in a single semiconductor quantum dot provides an excellent system for testing fundamental aspects of quantum dynamics and coherence. The central quantity characterizing the response of an electron or hole spin to an applied magnetic field, the Land\'e $g$ factor, has been measured optically\cite{optical-g-dot0,optical-g-dot1,optical-g-dot2,optical-g-dot3} and electrically\cite{elec-g-dot1,elec-g-dot2,Tarucha,Kouwenhoven}. For large dots\cite{Tarucha,Kouwenhoven,Valin-Rodriguez-2004,Rossler-holes}, such as those defined lithographically\cite{Tarucha} or by electrical gates\cite{Kouwenhoven}, the magnetic fields of interest are usually large enough that the magnetic length is smaller than the dot diameter. In this limit the $g$ factors are closely related to those of quantum wells, and theory\cite{Valin-Rodriguez-2004} appears to agree with experiment\cite{Kouwenhoven}. The theoretical situation is much less satisfactory for small dots --- asymmetric structures grown by self assembly in the molecular beam epitaxy (MBE) growth process or spherical nanocrystals grown by chemical synthesis.  Although several phenomena known to affect $g$-factors in quantum wells\cite{Ivchenkobook,Winklerbook} have been explored in quantum dots\cite{Efros-dots,Kiselev-elec-dots,Whaley-dots,GeSiQD,Prado-dots,Prado2004}, 
the electronic states in dots are discrete, and thus differ qualitatively from semiconductors with unbounded motion in one or more directions\cite{Bimberg}.  Many dot properties, such as the sharply-peaked optical transitions, thus resemble those of atoms more than bulk semiconductors. If quantum dots are considered as ``artificial atoms'' and approached with techniques developed for magnetic atom dopants in solids\cite{VanVleck}, then the relevant quantity is the ratio of the energy splitting between different angular momentum states to the spin-orbit interaction, and for a strong confining potential the $g$ factor of dots should approach $2$. 

\begin{figure}
\includegraphics[width=0.9\columnwidth]{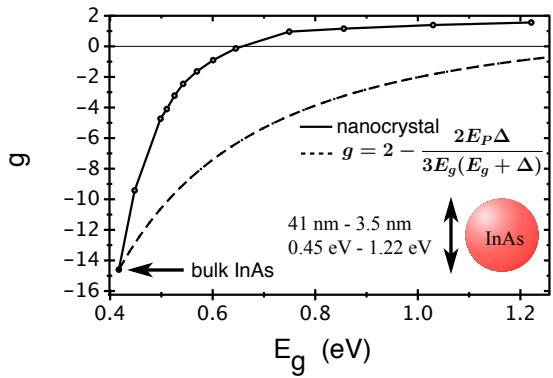}
\caption{ InAs nanocrystal electron $g$ factor as a function of dot size (parameterized by $E_g$).  The effects of angular momentum quenching are clearly seen as $g \rightarrow 2 $ for smaller nanocrystals.
 $10~ \mathrm{eV}$ barriers were used to exclude the wavefunction from the barriers. The dashed line shows the $g$ factor obtained from the bulk formula, using $E_g$ for the nanocrystal (including the confinement energy).}
\label{spherical}
\end{figure}

\begin{figure}
\includegraphics[width=0.9\columnwidth]{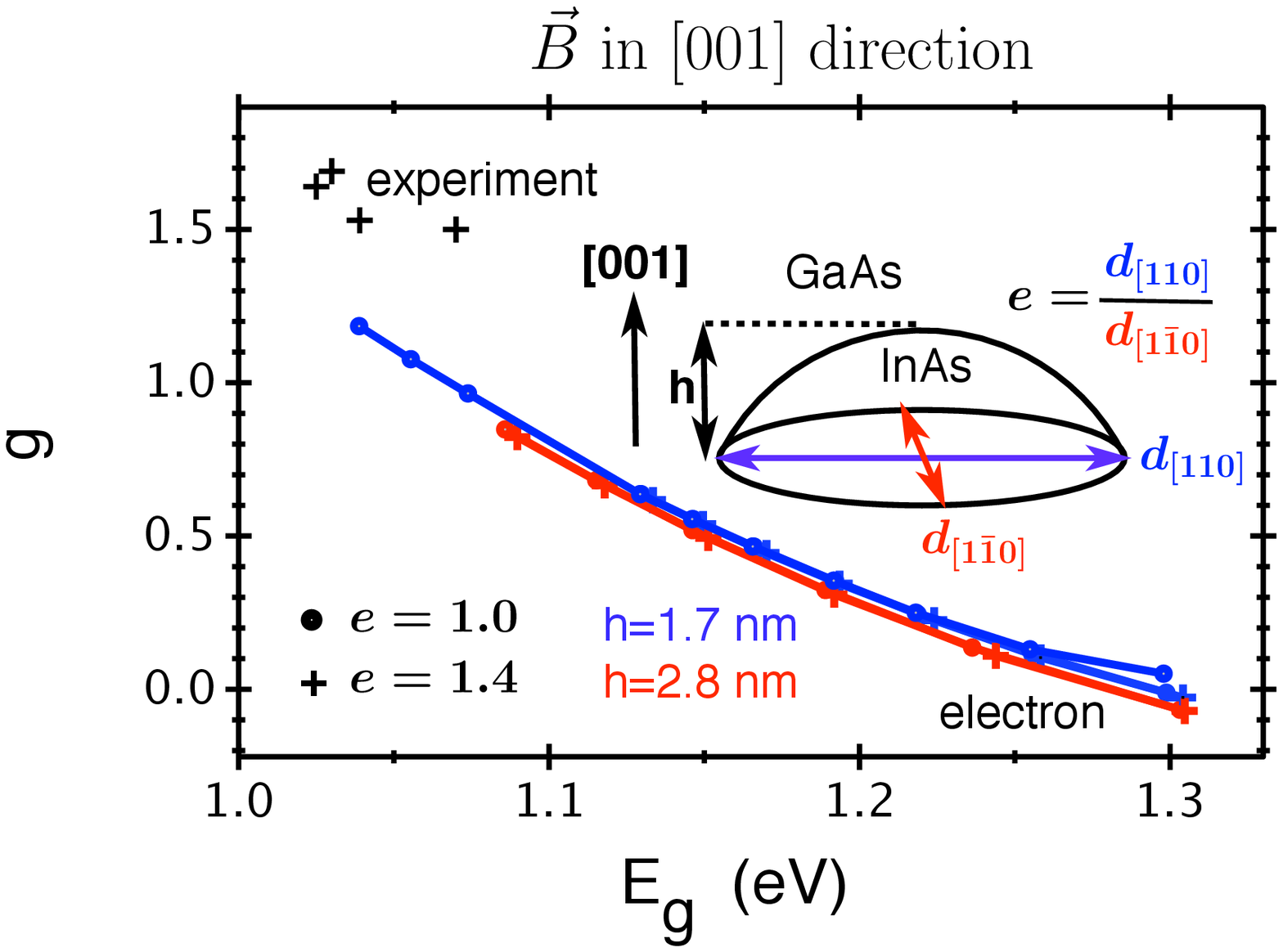}
\includegraphics[width=0.9\columnwidth]{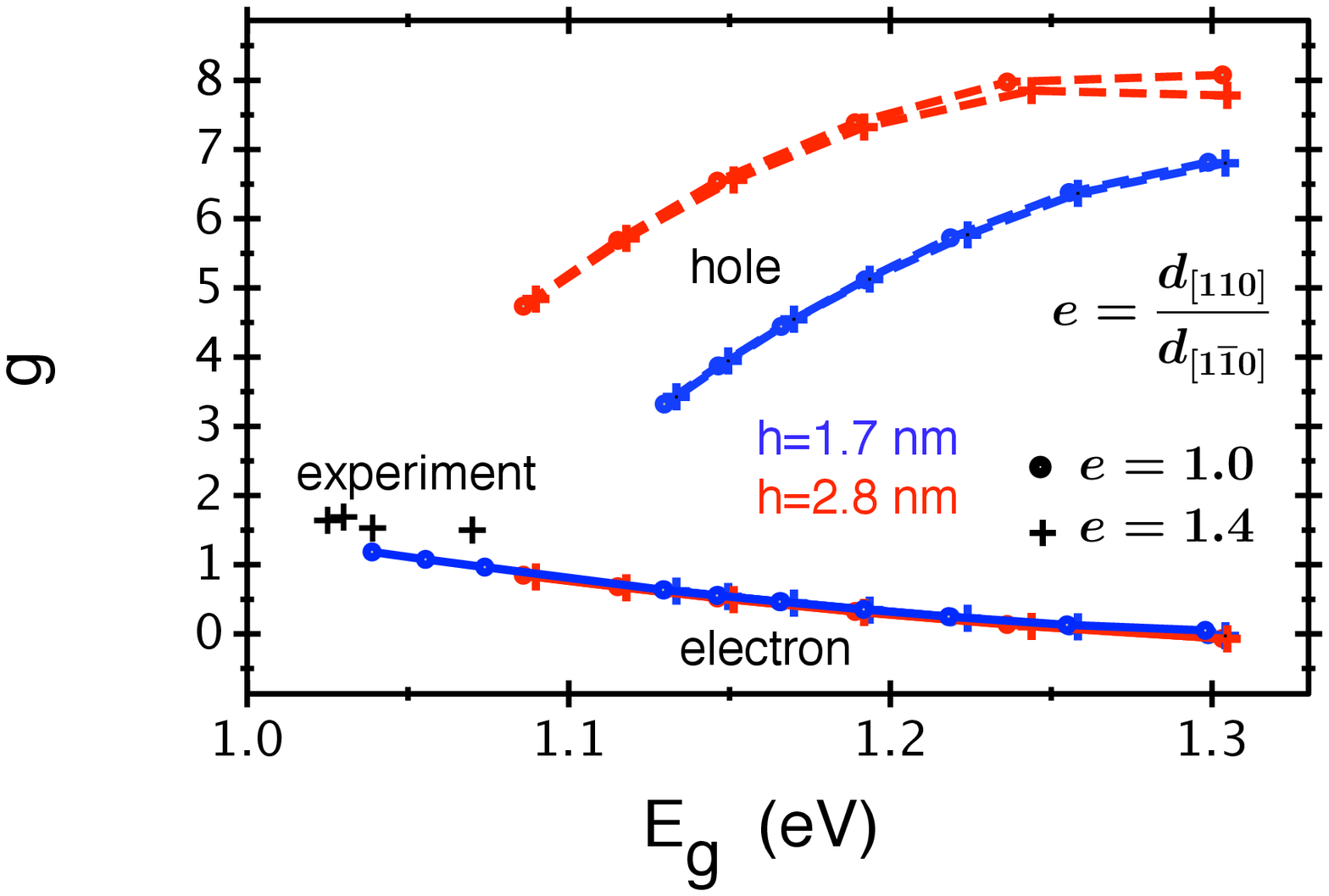}
\caption{InAs/GaAs quantum dot $g$ factors for $\vec B$ in the $[001]$ direction as a function of the dot size (parameterized by $E_g$), for various dot shapes.  
Solid lines are for the lowest energy electron, dashed lines are for the highest energy hole.
Experimental values are from Ref.  \cite{elec-g-dot1}
}\label{g-001}
\end{figure}

\begin{figure}
\includegraphics[width=0.9\columnwidth]{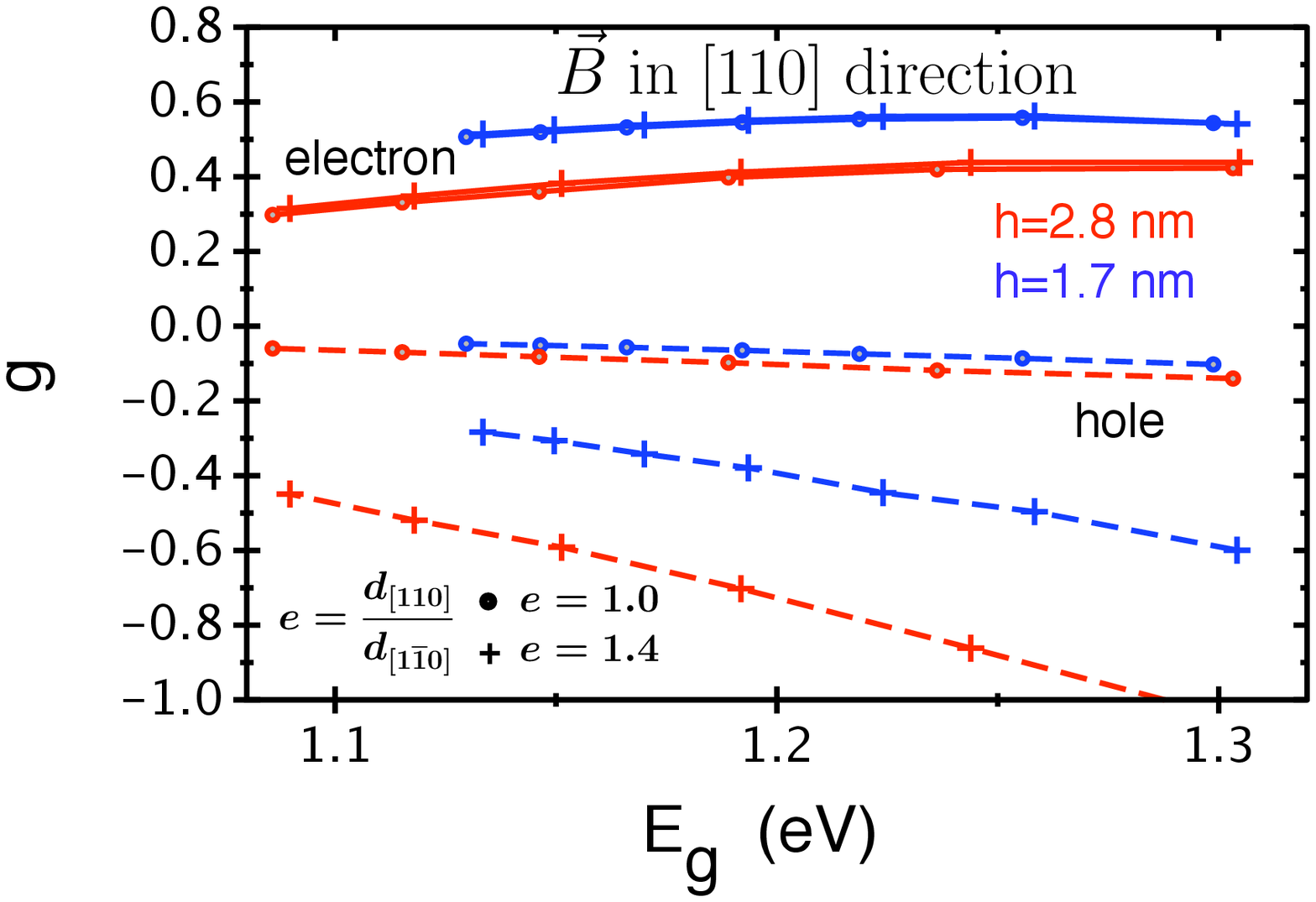}
\includegraphics[width=0.9\columnwidth]{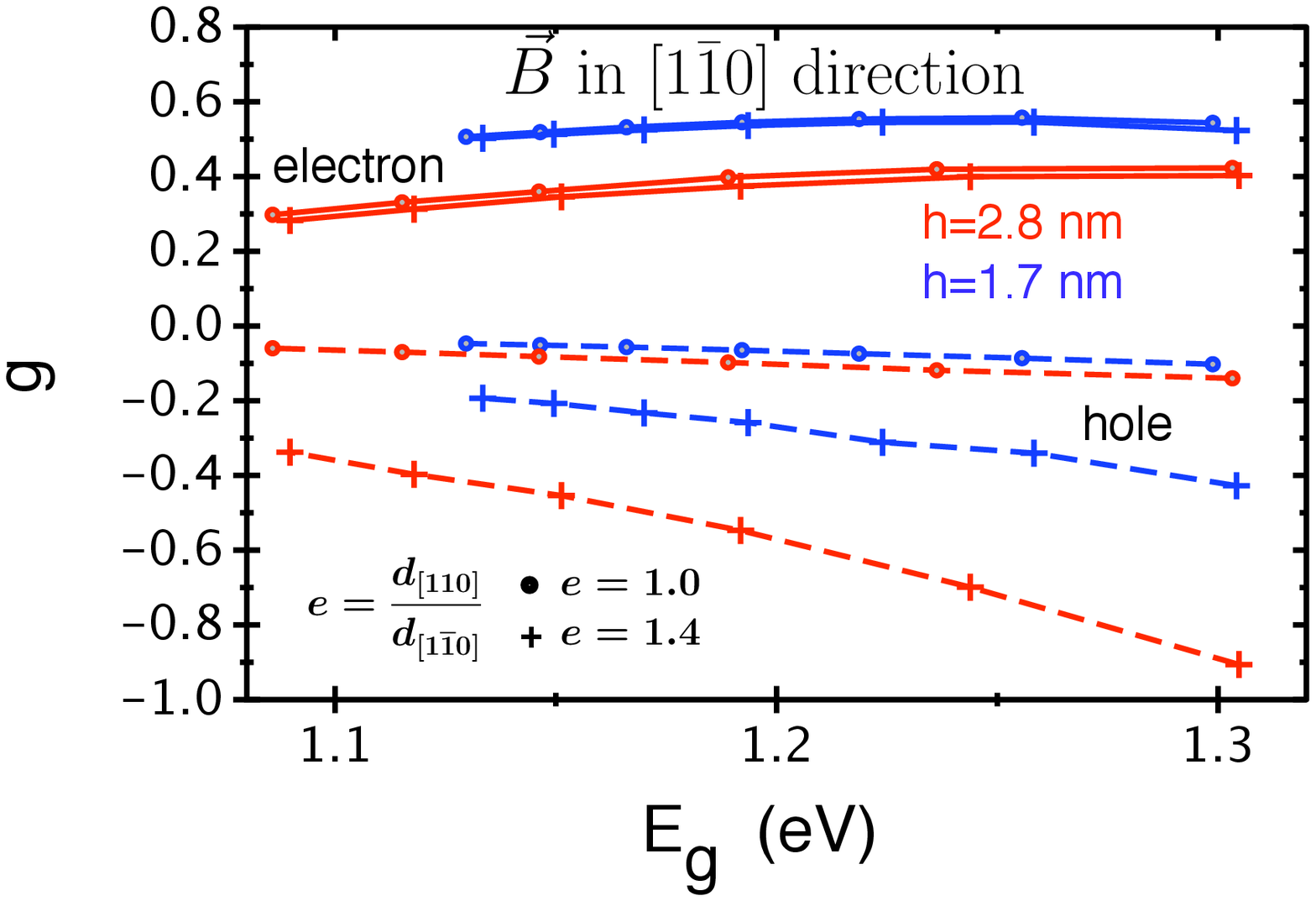}
\caption{InAs/GaAs quantum dot  $g$ factors along the $[110]$ and $[1\overline{1}0]$ directions
 as a function of dot size (parameterized by $E_g$).
Solid lines are for the lowest energy electron, dashed lines are for the highest energy hole.}\label{g-inplane}
\end{figure}

Here we show that  the $g$ factor of an electron or a hole in a quantum dot depends significantly on an atom-like property:  the quenching of orbital angular momentum through quantum confinement.  To identify this effect we must consider also the known bulk-like effects on the $g$ factors from dot strain and composition.  These include\cite{Efros-dots,Kiselev-elec-dots,Whaley-dots,GeSiQD,Prado-dots,Prado2004} the modification of the electron or hole ground-state energy in the dot, the relative proportion of dot or barrier material the wavefunction occupies, and valence band mixing (typically of heavy and light states).  In bulk semiconductors the conduction-band $g$ factor is\cite{ma-roth-lax},
\begin{eqnarray}
g = 2 - \frac{2 E_P \Delta}{3 E_g (E_g+\Delta)},
\label{rotheqn}
\end{eqnarray}
where $E_P$ is the Kane energy element, $E_g$ is the band gap, and $\Delta$ is the spin orbit coupling.  
For unstrained spherical InAs nanocrystals with hard wall boundary conditions only $E_g$ changes, yet Eq.~(\ref{rotheqn}) is a very poor predictor of $g$ factors in these dots (Fig.~\ref{spherical}).  For MBE-grown InAs/GaAs dots we find that the electron $g$ factor predicted from Eq.~(\ref{rotheqn}) for the strained InAs is negative, as is the known $g$ factor for unstrained GaAs. Yet the electron $g$ factors for such dots are positive over almost the entire size range (Fig.~\ref{g-001}). Thus the bulk-like approach to $g$ factors in these quantum dots, averaging the $g$ factors over the dot and barrier material\cite{Efros-dots,Kiselev-elec-dots}, also fails. The competing influence of atom-like and bulk-like effects predict that the growth-direction electron $g$ factor increases with increasing dot size, whereas considering only bulk-like effects leads to the opposite result. Our results agree with recent experiments on electron $g$ factors\cite{elec-g-dot1}. We predict hole $g$ factors have large positive values for magnetic fields in the growth direction, and small negative values in the in-plane directions(Fig.~\ref{g-001}-\ref{g-inplane}), and are sensitive to dot shape.  The calculations presented here are for  InAs spherical nanocrystals and  InAs/GaAs MBE-grown quantum dots, but our qualitative conclusion --- that the ``artificial atom'' viewpoint is vitally important to understanding $g$ factors --- applies to all small quantum dots. 


Our results also point the way towards electric-field control of $g$ factors in quantum dots.  $g$ factor control via manipulation of the electronic wavefunction in quantum wells has already been used to control spin precession\cite{Salis-g factor}, and to drive spin resonance\cite{Levy-g-TMR}. Due to the large size of quantum dots compared to atoms, moderate voltages applied by electrical gates\cite{Bimberg} can modify the dot shape, energy levels, and $g$, and thus drive spin resonance in a static magnetic field. Such control of electron spin resonance in an individual quantum dot could assist ultrafast manipulation of information encoded in electron or hole spin, as well as permit single-qubit gate operations for quantum computation\cite{DiVincenzo-Loss-1998}.

We have calculated quantum dot $g$ factors by calculating the spin splittings in a magnetic field of $|B| = 0.1 ~T$.
The sign of $g$ was determined by examination of the wave functions to see if the spin of the lower energy state was parallel or anti-parallel to $\vec B$.
The calculations were performed using 8-band strain dependent ${\bf k}\cdot {\bf p}$ theory in the envelope approximation  with finite differences on a real space grid\cite{Pryor.prb.1998}. Material parameters were taken from Ref.~\onlinecite{Vurgaftman} assuming $T=0$K.

The magnetic field was included by coupling to both the envelope function and the electron spin.
The  envelope was coupled to $\vec B$ by  making all difference operators covariant using the standard prescription for introducing gauge fields on a lattice.  For example,
\begin{eqnarray}
\frac{\psi(\vec r+\epsilon \hat x) - \psi(\vec r-\epsilon \hat x)} {2 \epsilon } \rightarrow\\
 \frac{\psi(\vec r+\epsilon \hat x) U_x(\vec r)  - \psi(\vec r-\epsilon \hat x) U_x^\dagger ( \vec r-\epsilon \hat x ) } {2 \epsilon }
\end{eqnarray}
where $\epsilon$ is the grid spacing and $U_x(\vec r )$ is the phase acquired by an electron hopping from the site at $\vec r$ to the site at $\vec r+\epsilon \hat x$.
The $U$'s were determined by the requirement that transport around a plaquette produced the Aharonov-Bohm phase corresponding to the encircled flux, for example
\begin{eqnarray}
U_x( \vec r) U_y( \vec r+ \epsilon \hat x ) U_x^\dagger ( \vec r + \epsilon \hat y ) U_y^\dagger ( \vec r ) 
= \exp( i \epsilon^2 B_\perp e/ \hbar)
\end{eqnarray}
where $B_\perp$ is the magnetic field component perpendicular to the plaquette.
The electron spin was coupled to $\vec B$ though a Pauli term for the Bloch functions,  given by
\begin{eqnarray}
H_s = \frac{\mu_B}{2}  \vec B \cdot \begin{pmatrix} 2\vec \sigma  & 0& 0\\ 0 &\frac{4}{3}\vec J & 0\\ 0
& 0 & \frac{2}{3}\vec \sigma \end{pmatrix}
\end{eqnarray}
where $\mu_B$ is the Bohr magneton and $\vec \sigma$ and $\vec J$ are the spin matrices for spin $1/2$ and $3/2$ respectively.  The $g$ factors for the Bloch functions are $2$, $\frac{4}{3}$, and $\frac{2}{3}$ for the conduction, valence, and spin-orbit bands respectively.
The Bloch function $g$ factors are determined solely by the angular momentum of the Bloch states.

As these numerical calculations are performed in a finite box, numerical artifacts may arise if the total  magnetic flux through the box is not an integer number of the flux quantum corresponding to a single electron\cite{Pryor.prb.1991}.  
To avoid any such problems, the value of $\vec B$ was modified around the edges of the box (within the barrier material) so as to make the total flux in each of the x, y, and z directions an integer.
This allowed the use of both hard wall and periodic boundary conditions, and comparison between the two boundary conditions was used to establish that the box was sufficiently large to avoid finite-size artifacts.

One of the advantages of this approach is that, once we settle on a bulk band basis, we do not further truncate our quantum dot state basis.  
Because the spin-orbit interaction is positive in the bulk semiconductor constituents of the dot, it is energetically favorable for the spin and orbital angular momenta to be anti-parallel to each other.  
(This is the ultimate reason for $g<2$).
As many individual states make positive or negative contributions to $g$, a calculation in a truncated dot state basis runs the risk of an unbalanced choice of positive or negative contributing states. This can produce $g>2$, unphysical for electrons.

We now summarize the physical picture we have developed to understand the calculations shown in Figs.~\ref{spherical}-\ref{g-inplane}.  First consider the known origin of the effective $g$ for conduction electrons in a bulk semiconductor\cite{ma-roth-lax}.  When a magnetic field is applied, the orbital part of the wavefunction is modified into Landau levels, corresponding to quantized orbital angular momentum around the axis of the magnetic field.  The Zeeman energy now splits the lowest Landau level into two spin-polarized Landau levels, one with spin parallel to the quantized orbital angular momentum and one antiparallel.  Although the bare $g=2$ lowers the energy of the parallel spin state and raises that of the antiparallel state, the spin-orbit interaction preferentially aligns spin {\it antiparallel} to the orbital angular momentum. When that effect is absorbed into an effective $g$, it makes $g<2$.  The situation is modified significantly in quantum dots. Instead of beginning with a continuous spectrum, which then is modified into Landau levels, the spectrum in quantum dots (with or without a magnetic field) is discrete. Thus the modification of the lowest energy electron state (C1) pair is proportional to the ratio of the cyclotron energy to the energy splitting between that pair of states and the next lowest pair, $\hbar\omega_c/(E_{C2}-E_{C1})$. For all but the largest dots this ratio is very small, and the resulting orbital angular momentum of the C1 states in a magnetic field is very small, leading to $g\rightarrow 2$.  We refer to this effect as quenching of the orbital angular momentum, and note that there are similarities between this picture and that developed for crystal-field splitting of degenerate $d$ and $f$ states for paramagnetic impurities in insulating solids\cite{VanVleck} and for splitting of the heavy hole and light hole subband energies in quantum wells\cite{Winklerbook}. A significant difference, however, is that these previous cases involve the splitting of a discrete number of degenerate atomic-like states, whereas the quenching here is of a continuum to individual Kramers doublets.

Figure~\ref{spherical} shows our calculations for spherical InAs nanocrystals, and the bulk-like formula\cite{ma-roth-lax}. The calculations were done with a high barrier (10~eV) to avoid any leakage of the wavefunction outside the dot. For large dots ($E_g \sim 0.41$~eV) the two agree, however they rapidly diverge for smaller dots; when the confinement energy equals the bulk band gap of InAs the deviation from $g-2$ predicted by Eq.~(\ref{rotheqn}) is six times larger than the actual value. Hole $g$ factors for these nanocrystals show similar evidence of orbital angular momentum quenching (not shown). 
Note that the quenching of orbital angular momentum for the electron and hole states  is compatible with the high fidelity selection rules for generating spin-polarized carriers in dots with optical means\cite{PryorFlattePRL}, because the optical transitions connect states with specific angular momentum (valence and conduction) whereas the $g$ factors probe how much angular momentum admixture is possible for $B\ne 0$.

Fig.~\ref{g-001} shows $g$ factors for the lowest-energy electron state and hole state for spherical-cap InAs/GaAs dots as the dot size decreases (and $E_g$ increases). The dots have a height $h$ in the growth direction, [001], and may be circular or elliptical (extended in the [110] direction), according to the ratio $e=d_{[110]}/d_{[1\overline{1}0]}$. In Fig.~\ref{g-001}, $h$ and $e$ are fixed as the size changes (dots with the same $h$ and $e$ and different $E_g$ have different size base lengths $d_{[110]}$ and $d_{[1\overline{1}0]}$).  For both the electrons and holes the deviation of $g$ from the quenched case ($g=2$) increases as $E_g$ increases (and the dot size decreases).  
For smaller dots, excited bound states become squeezed out of the dot, and the states with which the ground state mixes are increasingly from the continuum of states in the barrier material.
Hence in the limit of a very small dot  $g$ approaches the value corresponding to the bulk barrier material.
For very large dots, $g$ should correspond to the strained material within the dot.  However, this limit requires an extremely large dot, and is not reached for the computationally tractable dot sizes considered here (it is in the nanocrystal calculations of Fig.~\ref{spherical}). Measurements for large dots\cite{elec-g-dot1} are shown as well. These measurements cannot determine the sign of the $g$ factor; we identify the sign to be positive based on our calculations.

Using Eq.~\ref{rotheqn} with $E_g$ and $\Delta$ replaced by the values calculated for bulk InAs with the same strain as the InAs in the dot leads to the electron $g \approx -6$.  
Increasing $E_g$ to include the largest possible confinement energy for a dot state gives $g \approx -0.5$.  The GaAs barrier has $g=-0.44$, so any averaging approach would produce a negative $g$ factor, and the $g$ factor should decrease as the quantum dot size increases. Due to orbital angular momentum quenching the actual electron $g$ factors shown in Fig.~\ref{g-001} are positive, except for the largest dots considered, which have $g\approx -0.05$, and the $g$ factor increases as the quantum dot size increases --- the opposite trend.

As the appproximate dot shape is frequently unknown, the $[001]$ hole $g$ factors (Fig.~\ref{g-001}) or in-plane electron $g$ factors  (Fig.~\ref{g-inplane}) may be usable as diagnostics for the dot height, and the in-plane hole $g$ factors (Fig.~\ref{g-inplane}) for the elongation. Use of in-plane electron $g$ factors to diagnose dot height may be particularly effective because their dependence on $E_g$ and $e$ is weak.   In-plane hole $g$ factors, however, depend strongly on both the height and elongation, but not on $E_g$.   The in-plane hole $g$ factors are much smaller in magnitude than the hole $g$ factors along the growth direction.  This is qualitatively  similar to the results of Ref.~\onlinecite{GeSiQD} for Si/Ge quantum dots. However, for InAs/GaAs quantum dots  the sign of the hole $g$ factor differs for  the growth direction and in-plane.

We have shown that the $g$ factors in quantum dots are dominated by the atom-like phenomenon of angular momentum quenching, leading to electron and hole $g$ factor values and trends that differ substantially from those seen in bulk semiconductors, and which are much closer to the bare value of 2.  For InAs/GaAs dots, strain and geometric asymmetry also contribute to $g$, and result in strong $g$ factor anisotropy, especially for hole states. Careful study of the anisotropy of the hole $g$ factors can provide approximate information about the dot shape. Our unexpected results for the electron and hole $g$ factors suggest that the "artificial atom" picture promises additional surprises for the physics of spin dynamics in quantum dots.

This work was supported by DARPA/ARO DAAD19-01-1-0490.

\end{document}